# Degree of entanglement for two qubits

Ayman F. Abouraddy, Bahaa E. A. Saleh[†], Alexander V. Sergienko,

and Malvin C. Teich

*Quantum Imaging Laboratory,[‡] Departments of Electrical & Computer Engineering and*

*Physics, Boston University, 8 Saint Mary's Street, Boston, MA 02215-2421*

We demonstrate that any pure bipartite state of two qubits may be decomposed into a superposition of a maximally entangled state and an orthogonal factorizable one. Although there are many such decompositions, the weights of the two superposed states are, remarkably, unique. We propose a measure of entanglement based on this decomposition. We also demonstrate that this measure is connected to three measures of entanglement previously set forth: maximal violation of Bell's inequality, concurrence, and two-particle visibility.

*I. Introduction*

Entanglement is the quintessential property of quantum mechanics that sets it apart from any classical physical theory [1], and it is essential to quantify it in order to assess the performance of applications of quantum information processing [2]. An entangled state is a multi-particle state that cannot be factored into states of the individual particles. Many measures of entanglement proposed in the past have relied on either the Schmidt decomposition [3] or decomposition in a 'magic basis' [4]. In this paper we devise a new measure of entanglement for pure bipartite states of two qubits. Our definition is based on a decomposition of the state vector as a superposition of a maximally entangled and a factorizable state vector. We discuss the connection between our definition of the degree of entanglement and several related concepts previously discussed in the literature, and demonstrate that these seemingly unconnected concepts are actually identical.

---

[†] Electronic address: besaleh@bu.edu

[‡] URL: http://www.bu.edu/qil

## *II. Definitions*

A bipartite state $|\Psi_f\rangle \in H = H_1 \otimes H_2$, is said to be factorizable if it can be factored into a product, $|\Psi_f\rangle = |\Psi_1\rangle \otimes |\Psi_2\rangle$, where $|\Psi_1\rangle \in H_1, |\Psi_2\rangle \in H_2$, and $H_1$ and $H_2$ are the Hilbert spaces of the individual particles. An entangled state is one for which this is not possible. A maximally entangled bipartite state $|\Psi_e\rangle$ satisfies the conditions $\text{Tr}_1(|\Psi_e\rangle\langle\Psi_e|) = \frac{1}{2}\mathbf{I}_2$ and $\text{Tr}_2(|\Psi_e\rangle\langle\Psi_e|) = \frac{1}{2}\mathbf{I}_1$, where $\text{Tr}_1$ and $\text{Tr}_2$ stand for tracing over the subspaces $H_1$ and $H_2$, respectively, and $\mathbf{I}_1$ and $\mathbf{I}_2$ are the identity operators in $H_1$ and $H_2$, respectively. This implies that each particle, when considered alone, is in a maximally mixed state, although the state of the system as a whole is pure.

## *III. Degree of Entanglement*

For Hilbert spaces $H_1$ and $H_2$ of dimension 2, i.e., when each particle is a qubit, the general bipartite state may be expanded in the $\{|0\rangle, |1\rangle\}$ bases of $H_1$ and $H_2$ in the usual form

$$|\Psi\rangle = \alpha_1|00\rangle + \alpha_2|01\rangle + \alpha_3|10\rangle + \alpha_4|11\rangle, \tag{1}$$

where $\sum_j |\alpha_j|^2 = 1$. The state may also be written in terms of a Schmidt decomposition [3],

$$|\Psi\rangle = \kappa_1|x_1, y_1\rangle + \kappa_2|x_2, y_2\rangle, \tag{2}$$

where $\{|x_1\rangle, |x_2\rangle\}$ and $\{|y_1\rangle, |y_2\rangle\}$ are orthonormal bases for $H_1$ and $H_2$, respectively, and $\kappa_1$ and $\kappa_2$ are real and non-negative coefficients satisfying $\kappa_1^2 + \kappa_2^2 = 1$ and $\kappa_1 \geq \kappa_2$.

We propose a *different decomposition* that will lead to a definition for the degree of entanglement:

$$|\Psi\rangle = p|\Psi_e\rangle + \sqrt{1-p^2}\, e^{i\varphi}|\Psi_f\rangle. \tag{3}$$

Here $|\Psi_e\rangle$ is a maximally entangled normed state, $|\Psi_f\rangle$ is a factorizable normed state orthogonal to $|\Psi_e\rangle$ ($\langle\Psi_e|\Psi_f\rangle = 0$), and $p$ and $\varphi$ are real numbers.

It is shown in the Appendix that this decomposition *always exists and is not unique, but the parameter $p$ is unique*. This is a quite remarkable result and, to the best of our knowledge, has not been observed before in the literature on entanglement measures.

An entire family of $\{|\Psi_e\rangle, |\Psi_f\rangle\}$ pairs exists for each state, but all have the same value of $p$. It remains to demonstrate how this family may be generated for a given state. It is shown in the Appendix that the decomposition in Eq. (3) may be obtained from the (unique) Schmidt decomposition given in Eq. (2) by a local unitary transformation, $\mathbf{U} = \mathbf{U}_1 \otimes \mathbf{U}_2$, where

$$\mathbf{U}_1 = \begin{bmatrix} a & -be^{i\theta} \\ be^{-i\theta} & a \end{bmatrix}, \quad \mathbf{U}_2 = \begin{bmatrix} b & -ae^{-i\theta} \\ ae^{i\theta} & b \end{bmatrix}, \tag{4}$$

with $a$ and $b$ positive real numbers, $a^2 + b^2 = 1$, and $\theta$ an arbitrary phase. Applying this unitary transformation to Eq. (2) with $a = \sqrt{\dfrac{\kappa_1}{\kappa_1 + \kappa_2}}$, $b = \sqrt{\dfrac{\kappa_2}{\kappa_1 + \kappa_2}}$, gives

$$|\Psi\rangle = \frac{p}{\sqrt{2}}\left(|u_1, v_1\rangle + |u_2, v_2\rangle\right) + \sqrt{1-p^2}\, e^{i\theta} |u_1, v_2\rangle, \tag{5}$$

which is of the form of Eq. (3). Reversing the values of $a$ and $b$ gives

$$|\Psi\rangle = \frac{p}{\sqrt{2}}\left(|u_1, v_1\rangle + |u_2, v_2\rangle\right) + \sqrt{1-p^2}\, e^{i\theta} |u_2, v_1\rangle \tag{6}$$

which, again, is of the form of Eq. (3). The parameters $a$ and $b$ are unique, whereas $\theta$ is a free parameter.

As an example, the state

$$|\Psi\rangle = \frac{1}{\sqrt{3}}\left(|00\rangle + |01\rangle + |11\rangle\right) \tag{7}$$

may be decomposed in the form of Eq. (3) with $|\Psi_e\rangle = \dfrac{1}{\sqrt{2}}(|00\rangle + |11\rangle)$, $|\Psi_f\rangle = |01\rangle$, and $p = \sqrt{\dfrac{2}{3}}$. Another decomposition can make use of the states

$|\Psi_e\rangle = \dfrac{\sqrt{2}}{10}(3|00\rangle + 4|01\rangle - 4|10\rangle + 3|11\rangle)$ and $|\Psi_f\rangle = \dfrac{1}{5}(2|00\rangle + |01\rangle + 4|10\rangle + 2|11\rangle)$, with the

same value of $p = \sqrt{\dfrac{2}{3}}$. It can be easily demonstrated that using $a = \sqrt{\dfrac{5+\sqrt{5}}{10}}$, $b = \sqrt{\dfrac{5-\sqrt{5}}{10}}$, and $\theta = 0$ in $\mathbf{U}_1$ and $\mathbf{U}_2$ in Eq. (4) leads to the first decomposition whereas using $\theta = \pi$ leads to the second.

Now that we have established that a state may be decomposed into a superposition of maximally entangled and factorizable parts, it is natural to use the squared weight $p^2$ as a measure of the *degree of entanglement* $P_E \equiv p^2$. This new measure $P_E$ is bounded by $0 \leq P_E \leq 1$ and is invariant under local unitary transformations. It is clear from the Appendix that the state defined in Eqs. (1) and (2) has a degree of entanglement

$$P_E = p^2 = 2|\alpha_1\alpha_4 - \alpha_2\alpha_3| = 2\kappa_1\kappa_2. \tag{8}$$

We may justify using $P_E$ as a measure of the degree of entanglement in another way. Bell's inequality [5] tests the nonlocality of quantum mechanics that was challenged by Einstein, Podolsky, and Rosen (EPR) [6]. The form of Bell's inequality that has principally been put to the test is that due to Clauser, Horne, Shimony, and Holt (CHSH) [7]. Their formulation requires evaluating the following quantity for a bipartite state:

$$f(\Psi) = |E(c,d) + E(c',d) + E(c,d') - E(c',d')| \leq 2, \tag{9}$$

where $c$ and $c'$ are two observables of the first particle and $d$ and $d'$ are two of the second, such that they all have a maximum absolute expected value of 1, and $E(c,d)$ is the expected value of the correlation of $c$ and $d$, and so on for the other expected values. Local physical theories satisfy this inequality whereas quantum mechanics violates it for a judicious choice of measurements. It has been shown [8] that the maximum violation of this inequality is $f_{\max}(\Psi) = 2\sqrt{1 + 4\kappa_1^2\kappa_2^2}$. Using Eq. (8), this relationship can be rewritten as $f_{\max}(\Psi) = 2\sqrt{1 + P_E^2}$, i.e., the maximum violation of the CHSH form of Bell's inequality for a pure state is limited by $P_E$.

## IV. Previous Measures of Entanglement

A. Wootters' measure

In Ref. [4] Wootters presents a measure of the entanglement of a bipartite state of 2 qubits that is denoted the *concurrence*, $C$. The entanglement of formation of the bipartite state can always be formulated as a function of $C$, which is defined as follows. First write the state in the 'magic' basis: $|e_1\rangle = \frac{1}{\sqrt{2}}(|00\rangle + |11\rangle)$, $|e_2\rangle = \frac{i}{\sqrt{2}}(|00\rangle - |11\rangle)$, $|e_3\rangle = \frac{i}{\sqrt{2}}(|01\rangle + |10\rangle)$, $|e_4\rangle = \frac{1}{\sqrt{2}}(|01\rangle - |10\rangle)$, such that $|\Psi\rangle = \sum_j \beta_j |e_j\rangle$, then $C = \left|\sum_j \beta_j^2\right|$. It is straightforward to show that $P_E$ is identical to $C$, thereby giving meaning to the concept of concurrence and demonstrating that the magic basis is unnecessary for arriving at this measure of entanglement [4].

B. Shimony's measure

In Ref. [9] Shimony defines the degree of entanglement, $E(\Psi)$, to be the minimum 'distance' between the state and any factorizable state, $E(\Psi) = \frac{1}{2}\min\||\Psi\rangle - |\Psi_f\rangle\|^2$, where the minimum is taken over the set of factorizable states. He shows, based on the Schmidt decomposition in Eq. (2), that $E(\Psi) = 1 - \kappa_1$. This definition suffers from the disadvantage of scaling from 0 to $0.293$, instead of the more satisfying range of 0 to 1, and also from the arbitrariness of the power 2 in the 'distance'. More germane, perhaps, is the fact that this is a measure of the distance to the set of factorizable states in the Hilbert space. Every normalized maximally entangled state has a projection of length $1/2$ onto the set of factorizable states (Eq. (2) with $\kappa_1 = \kappa_2 = 1/\sqrt{2}$). This definition thus differs conceptually from the definition we propose in Eq. (3), which projects the state simultaneously onto the set of maximally entangled and factorizable states.

C. Two-particle visibility

Consider a two-particle interferometer [10]. A two-particle source in an unknown pure state emits one particle in the $\{|x_1\rangle, |x_2\rangle\}$ basis and another in the $\{|y_1\rangle, |y_2\rangle\}$ basis. The particles encounter unitary transformations $\mathbf{U}_1$ and $\mathbf{U}_2$, which transform the bases to $\{|u_1\rangle, |u_2\rangle\}$ and $\{|v_1\rangle, |v_2\rangle\}$, respectively. Detectors register the singles rates, $P_1(u_1)$, $P_1(u_2)$, $P_2(v_1)$, $P_2(v_2)$, and the coincidence rates $P_{12}(u_1, v_1)$, $P_{12}(u_1, v_2)$, $P_{12}(u_2, v_1)$, $P_{12}(u_2, v_2)$. The aim is to define a two-particle visibility, $V_{12}$, that is representative of the degree of entanglement of the source and that is analogous, at least in its formal definition, to the visibility of classical interferograms [11].

Jaeger *et al.* [12,13] define a 'corrected' coincidence probability, $\overline{P}(u_1, v_1) = P_{12}(u_1, v_1) - P_1(u_1)P_2(v_1) + A$, where $A$ is a constant. They define the visibility as the ratio of the difference between the maximum and minimum values of $P_{12}(u_1, v_1)$ taken over all different $\mathbf{U}_1$ and $\mathbf{U}_2$, and the sum. However, the definition of $\overline{P}(u_1, v_1)$, as well as the choice of the value of $A$, is *ad hoc*.

In the conception presented here, the state at the output of $\mathbf{U}_1$ and $\mathbf{U}_2$ is written as $|\Psi\rangle = \alpha_1|u_1, v_1\rangle + \alpha_2|u_1, v_2\rangle + \alpha_3|u_2, v_1\rangle + \alpha_4|u_2, v_2\rangle$, so that

$$\overline{P}(u_1, v_1) = |\alpha_1\alpha_4|^2 - |\alpha_2\alpha_3|^2 + A = (|\alpha_1\alpha_4| - |\alpha_2\alpha_3|)(|\alpha_1\alpha_4| + |\alpha_2\alpha_3|) + A. \tag{10}$$

If we choose the phases of the elements of $\mathbf{U}_1$ and $\mathbf{U}_2$ such that $\sin\Phi = 0$, where $\Phi = \theta_1 + \theta_2 + \varphi_1 + \varphi_2$ (see the Appendix), then both $\alpha_1\alpha_4$ and $\alpha_2\alpha_3$ are real positive quantities and $\overline{P}(u_1, v_1) = A \pm P_E \dfrac{\alpha_1\alpha_4 + \alpha_2\alpha_3}{2}$; the quantity $\pm \dfrac{\alpha_1\alpha_4 + \alpha_2\alpha_3}{2}$ fluctuates as the parameters of $\mathbf{U}_1$ and $\mathbf{U}_2$ are changed. The value of $A$ should thus be chosen to be equal to the maximum absolute value of this latter quantity, which is $1/4$ when $|\alpha_1| = |\alpha_4|$ and $|\alpha_2| = |\alpha_3|$. One can show that the choice of $\mathbf{U}_1$ and $\mathbf{U}_2$ that leads to the above condition is the same one that leads to the results provided in Refs. [12] and [13], which were related to interferometric complementarities but not to the degree of entanglement. The authors in Ref. [13] found that $V_{12} = 2\kappa_1\kappa_2$, so that the measurement of two-particle visibility is tantamount to a measurement of the degree of entanglement $P_E$.

Note also that the visibilities of the singles rates (the one-particle visibilities) are all given by $\sqrt{1-P_E^2}$, so that in the context of our present construction, the complementarity of one- and two-particle visibilities [12, 13] follows immediately from the normalization of the state vector.

Another interesting conclusion emerges from the following considerations. The state $|\Psi_e\rangle$ offers no *welcher-weg* (which-way) information about the two particles since each particle considered separately is in a maximally mixed state, whereas $|\Psi_f\rangle$ provides definite *welcher-weg* information about the two particles. Thus the complementarity of one- and two-particle visibilities is the two-particle counterpart of the well-known complementarity for a single particle: that of *welcher-weg* information and interference visibility. In Ref. [13] the authors noted the similarity between these two complementarity relationships. The significance of this similarity is now clear.

We conclude that the proposed decomposition of Eq. (3) provides the underlying foundation for several seemingly different definitions of the degree of entanglement of a pure bipartite state of two qubits.

We are grateful to Zachary Walton for valuable discussions. This work was funded by the National Science Foundation and by the Center for Subsurface Sensing and Imaging Systems (CenSSIS), an NSF Engineering Research Center.

*Appendix: Properties of the new decomposition*

Apply the most general local unitary transformation $\mathbf{U} = \mathbf{U}_1 \otimes \mathbf{U}_2$ to the general bipartite state expressed in the Schmidt decomposition in Eq. (2):

$$\mathbf{U}_1 = \begin{bmatrix} a_1 & -a_2 \\ a_2^* & a_1^* \end{bmatrix}, \quad \mathbf{U}_2 = \begin{bmatrix} b_1 & -b_2 \\ b_2^* & b_1^* \end{bmatrix}, \tag{A1}$$

where $|a_1|^2 + |a_2|^2 = 1$ and $|b_1|^2 + |b_2|^2 = 1$; and $a_j = |a_j| e^{i\theta_j}$, $b_j = |b_j| e^{i\varphi_j}$, $j = 1, 2$; such that $|x_1\rangle \to a_1 |u_1\rangle + a_2^* |u_2\rangle$, and so on. After transformation, the state in Eq. (2) may then be written as

$$|\Psi\rangle = \beta_1 |u_1, v_1\rangle + \beta_2 |u_1, v_2\rangle + \beta_3 |u_2, v_1\rangle + \beta_4 |u_2, v_2\rangle, \tag{A2}$$

where $\beta_1 = \kappa_1 a_1 b_1 + \kappa_2 a_2 b_2$, $\beta_2 = \kappa_1 a_1 b_2^* - \kappa_2 a_2 b_1^*$, $\beta_3 = \kappa_1 a_2^* b_1 - \kappa_2 a_1^* b_2$, $\beta_4 = \kappa_1 a_2^* b_2^* + \kappa_2 a_1^* b_1^*$. If we impose the conditions $\beta_3 = 0$ and $\beta_1 = \beta_4$, we have $\kappa_2 |a_1||b_2| = \kappa_1 |a_2||b_1|$, $|a_1||b_1| = |a_2||b_2|$, $\theta_1 + \varphi_1 = \theta_2 + \varphi_2$. Solving the first two relationships, we obtain $|a_1| = |b_2| = \sqrt{\dfrac{\kappa_1}{\kappa_1 + \kappa_2}}$ and $|a_2| = |b_1| = \sqrt{\dfrac{\kappa_2}{\kappa_1 + \kappa_2}}$; we then have $\beta_1 = \beta_4 = \dfrac{p}{\sqrt{2}} e^{-i(\theta_1 + \varphi_1)}$ and $\beta_2 = \sqrt{1 - p^2} e^{i(\theta_1 - \varphi_2)}$, where $p^2 = 2\kappa_1 \kappa_2$. Since the Schmidt coefficients are unique for any given state, the parameter $p$ is also unique. We absorb the phases into the definition of $\mathbf{U}_1$ and $\mathbf{U}_2$ given in Eq. (A1) and thereby finally obtain the result given in Eq. (5). We can similarly impose the conditions $\beta_2 = 0$ and $\beta_1 = \beta_4$ in Eq. (A2) to obtain the result given in Eq. (6). A similar analysis, but used for a different purpose, is the starting point of Ref. [14].

The parameter $p$ may also be expressed in terms of the coefficients of $|\Psi\rangle$ in Eq. (1). A maximally entangled state takes the form $|\Psi_e\rangle = e^{i\gamma}\left(a_1|00\rangle + a_2|01\rangle - a_2^*|10\rangle + a_1^*|11\rangle\right)$, whereas a factorizable state takes the form $|\Psi_f\rangle = b_1|00\rangle + b_2|01\rangle + b_3|10\rangle + b_4|11\rangle$, where $\gamma$ is a phase, $|a_1|^2 + |a_2|^2 = \frac{1}{2}$, and $b_1 b_4 - b_2 b_3 = 0$. The coefficients of $|\Psi\rangle$ in Eq. (1) may be written in terms of the coefficients of $|\Psi_e\rangle$ and $|\Psi_f\rangle$, using Eq. (3), as $\alpha_1 = p e^{i\gamma} a_1 + \sqrt{1-p^2}\, e^{i\varphi} b_1$, and similarly for $\alpha_2$, $\alpha_3$, and $\alpha_4$. It readily follows that

$$\alpha_1 \alpha_4 - \alpha_2 \alpha_3 = \frac{1}{2} p^2 e^{i2\gamma} + p\sqrt{1-p^2}\, e^{i(\gamma+\varphi)}\left(a_1 b_4 + a_1^* b_1 - a_2 b_3 + a_2^* b_2\right). \tag{A3}$$

The expression in parentheses on the right hand side of Eq. (A3) is precisely the orthogonality condition $\langle \Psi_e | \Psi_f \rangle = 0$. It follows that $|\alpha_1 \alpha_4 - \alpha_2 \alpha_3| = \frac{1}{2} p^2$, completing the proof of Eq. (8).